\documentclass[twocolumn,preprintnumbers,amsmath,amssymb]{revtex4}

\usepackage{graphicx}
\usepackage{dcolumn}
\usepackage{bm}

\begin{document}


\title{Stochastic resonance in a single electron turnstile}

\author{Hisanao Akima}
 \email{akima@nakajima.riec.tohoku.ac.jp}
\author{Shigeo Sato}%
\author{Koji Nakajima}
\affiliation{%
Intelligent Nano-Integration System,
Research Institute of Electronical Communication,
Tohoku University,
2--1--1, Katahira, Aoba-ku, Sendai, 980--8577 Japan
}%

\date{\today}

\begin{abstract}
In this paper, we report stochastic resonance (SR) in a single electron turnstile. 
It has been known that SR emerges by the cooperation of a weak periodic signal and noise in a bistable system. 
A periodic signal produces switching between two stable states by lowering the potential barrier. 
Even if the amplitude of the signal is not large enough to make the potential barrier disappear, the switching can occur with the help of noise. 
As a result, an output signal-to-noise ratio (SNR), which shows how much the switching is synchronized with the signal, can be enhanced by increasing the noise. 
We have formulated the SR in a single electron turnstile into the adiabatic two-state theory of McNamara-Wiesenfeld and confirmed its manifestation by numerical simulation. 
These results could be applied to detection of the periodic motion of very small electric charge under noisy environment. 

\end{abstract}

\maketitle
\section{INTRODUCTION}
Stochastic resonance (SR) is a nonlinear phenomenon whereby the addition of noise can enhance the detection of weak stimuli. 
An optimal amount of added noise results in the maximum enhancement, whereas further increases in the noise intensity only degrade detectability. 
Although a large number of phenomena ranging from physics and engineering to biology and medicine have been studied (see the good review \cite{Moss}), the essential ingredients for SR consist of a threshold, subthreshold stimulus and noise in nonlinear systems. 
In this study, we especially consider a bistable system with a weak periodic signal and random noise. 
Experimental observation of SR in such bistable systems involved a Schmitt trigger circuit\cite{Fauve} and a radio frequency superconducting quantum interference device ~(rf-SQUID)~\cite{Rouse}~\cite{Hibbs}. 
An rf-SQUID utilizing SR could improve its performance by reducing its environmental noise sensitivity. 
It may be naturally expected that SR in a single electron circuit can be realized since the duality between an rf-SQUID and a single electron box (SEB) has been confirmed~\cite{Katsumoto}.
The SR in an SEB must be useful for detecting small charges under noisy environment. 
An rf-SQUID is composed of a Josephson junction, a superconducting loop inductance, and a bias current source in parallel. 
The number of flux quantums $n_f$ trapped in the loop is controlled by changing the bias current. 
On the other hand an SEB is composed of a tunnel junction, a gate capacitance and a bias voltage source in series. 
The number of excess electrons $n_e$ trapped in an island, which is the electrically isolated region between the tunnel junction and the gate capacitance, is controlled by changing the bias voltage. 
A perfect duality seems to exist, however, it is not true about bistability. 
While an rf-SQUID has bistability related to $n_f$ due to hysteresis property of a Josephson junction, an SEB does not have bistability related to $n_e$ because one tunnel junction does not show hysteresis property. 
Then let us consider a single electron turnstile~\cite{Geerligs} which is given bistability with four tunnel junctions in series. 
In the following sections, we formulate the SR in a single electron turnstile into the adiabatic two-state theory of McNamara-Wiesenfeld~\cite{McNamara} and confirm its manifestation by numerical simulation. 

\section{THEORY}
\subsection{Analysis of the dynamics}
\begin{figure}[tb]
\begin{center}
  \includegraphics{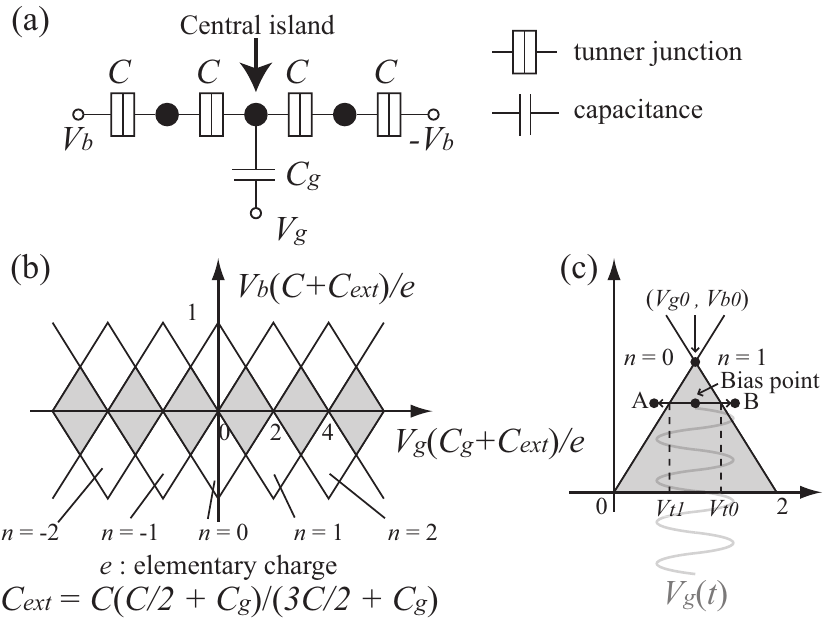}
  \caption{A single electron turnstile (a) and its stability diagram at zero temperature (b). The stability diagram enlarged near the bias point (c). An alternating gate voltage $V_{g}(t)$ swings the operating point.}
  \label{fig.turnstile}
\end{center}
\end{figure}

Figure \ref{fig.turnstile} shows the equivalent circuit of a single electron turnstile and its stability diagram. 
The stability diagram is a two-dimensional map of the stable states of a circuit~\cite{Wasshuber}. 
The x- and y-axes represent two bias voltages, namely the normalized gate voltage and the normalized bias voltage, respectively. 
At the operating points surrounded by each diamond in Fig.\ref{fig.turnstile}(b), the change of $n$, which signifies the number of excess electrons in the central island, leads to increase of free energy. 
Then electron tunneling is inhibited (Coulomb blockade~\cite{Averin}) and the circuit remains stable states. 
Thus the gray regions in which two neighboring diamonds are overlapped are bistable regions: two different stable states with different $n$ have been realized exclusively. 
If the gate voltage $V_{g}(t)=A_v\sin(2\pi f_st)+V_{g0}$ takes the operating point cut across the bistable region periodically like A$\to$ B$\to$ A $\to \cdots$ in Fig.\ref{fig.turnstile}(c), $n$ varies $0\to1\to0\to \cdots$.
Consequently such single electron transportation synchronized with the frequency $f_s$ can be realized and then current $I_s=ef_s$ flows. 
This property can be utilized for a current standard. 
Notice that the periodic motion of an electric charge $Q_s(t)=A_q\sin(2\pi f_st)$ polarized in the central island corresponds to the ac-component $V_s(t)=A_v\sin(2\pi f_st)$, where $A_v=A_q/C_g$ and $C_g$ is a gate capacitance. 
While the periodic motion of an outside electric charge near the central island produces the synchronized current flow in the same manner even without the ac-component of $V_g$.
Therefore a single electron turnstile can be used as a detector of the periodic motion of electric charge. 
Let us consider $Q_s(t)$ and $n(t)$ as input and output signals, respectively. 
Bistability related to $n(t)$ leads to SR with the help of some external noises such as fluctuation of voltages, fluctuation of background charge motion. 

Let us suppose the circuit is biased at the point ($V_{g0}, V_{b0}-\varepsilon$) as shown in Fig.\ref{fig.turnstile}(c), 
where $V_{g0}=e/(C_g+C_{ext})$, $V_{b0}=e/2(C+C_{ext})$, $e$ is elementary charge, $C$ is a tunnel capacitance, $C_{ext}\equiv C(C/2+C_g)/(3C/2 + C_g)$ and $\varepsilon$ is a small positive constant. 
The input charge signal $Q_s(t)$ forces the operating point of the circuit to fluctuate with the amplitude $A_q/C_g$ and the frequency $f_s$ around the bias point. 
We assume that Gaussian charge noise $Q_N(t)$ with cutoff frequency $f_c$ is added to $Q_s(t)$. 
Then the input gate voltage is given as $V_g(t)=V_{g0}+\{Q_s(t)+Q_N(t)\}/C_g\equiv V_{g0}+V_s(t)+V_N(t)$. 
Switching of $n(t)$ occurs at $V_g=V_{t0} (n=0\to 1)$ and $V_g=V_{t1} (n=1\to 0)$, 
where
\begin{eqnarray}
V_{t0}&=&2V_{g0}-2V_b(C+C_{ext})/(C_g+C_{ext}), \nonumber \\
V_{t1}&=&2V_b(C+C_{ext})/(C_g+C_{ext}),
\end{eqnarray}
and $V_b$ is a bias voltage. 
Note that the signal frequency $f_s$ has to meet the ``adiabatic limit" $1/f_s\gg \tau_t$, where $\tau_t$ is the tunneling time of electrons, in order to get the circuit settle in equilibrium states $n=$0 or 1. 
If tunneling time is negligibly small, the change of excess electrons accompany with tunneling at zero temperature is given as follows, 
\begin{eqnarray}
n(t_+)&=&\frac{1}{2}\Bigl\{ sgn(z) + 1\Bigr\}, \nonumber \\
z&\equiv&V_g(t_+) -[ n(t_-)V_{t1} + \{1-n(t_-)\}V_{t0}] \nonumber \\
&=&(V_{t0} - V_{t1}) n(t_-) - V_{t0} + V_g(t_+), \label{eq.n}
\end{eqnarray}
where $sgn(z)=1$ if $z>0$ and $sgn(z)=-1$ if $z<0$.
Subscripts $_+$ and $_-$ for $t$ mean after and before tunneling, respectively. 
The behavior defined by Eq.(\ref{eq.n}) is similar to that of a Schmitt trigger circuit. 
Hereafter we assume that operating temperature $T$ is not zero and then switching occurs stochastically in the vicinity of $V_{t0}$ and $V_{t1}$. 
Hence $n(t)$ should be treated as a continuous variable so that it indicates an expected value. 
Following the procedure derived for the Schmitt trigger circuit~\cite{McNamara}, the time evolution of $n(t)$ decaying with a time constant $\tau_t$ is given as follows, 
\begin{eqnarray}
\frac{dn}{dt} = -\tau_t^{-1} \Bigl[ n - \frac{1}{2}\{ \tanh(\beta z) + 1\}\Bigr], \label{eq.n_div}
\end{eqnarray}
where $\beta$ is the gain parameter related to $T$. 
Integrating the right side of Eq.(\ref{eq.n_div}) produces the potential $U(n,t)$ which governs the dynamics of $n(t)$. 
\begin{eqnarray}
U(n,t)&=&\tau_t^{-1} \Bigl[ \frac{1}{2}n(n-1) - \frac{1}{2\beta(V_{t0}-V_{t1})} \nonumber \\
&\times&\ln\cosh[\beta\{ (V_{t0}-V_{t1})n - V_{t0} + V_g(t)\}]\Bigr]. \label{eq.potential}
\end{eqnarray}
The signal $V_s(t)=Q_s(t)/C_g$ modulates $U(n,t)$ periodically and lowers the potential barrier between the two stable states $n=0$ and 1. 
Even if the amplitude of the signal is not large enough to make the potential barrier disappear (subthreshold signal), switching can occur with the help of the noise $V_N(t)=Q_N(t)/C_g$; this is the source of SR. 

\subsection{Stochastic resonance -- adiabatic two-state theory}
Since the dynamics of a single electron turnstile is described with the double-well potential $U(n,t)$, we can apply the adiabatic two-state theory~\cite{McNamara} in which SR in bistable systems have been studied. 
An output signal-to-noise ratio (SNR) shows how much the switching between two stable states is synchronized with the input signal. 
The addition of noise degrades the SNR in linear systems, while added noise results in the enhancement of the SNR in nonlinear systems in which SR occurs. 
Therefore the SNR versus input noise profile is sometimes taken to be the hallmark of SR. 
The SNR is computed by integrating the power spectral density (PSD) over the peak centered at the signal frequency $f_s$, and dividing by the mean noise power around $f_s$ and is found in dB~\cite{McNamara}, 
\begin{eqnarray}
\mathrm{SNR}=10 \log \Bigl[\frac{SG+N\Delta}{N\Delta}\Bigr]. \label{eq.SNR}
\end{eqnarray}
$G$ is the processing gain given as follows, 
\begin{eqnarray}
G=\frac{1}{N_{FFT}}\frac{[\sum_{i}^{N_{FFT}}w_i]^2}{\sum_{i}^{N_{FFT}}w_i^2},
\end{eqnarray}
where $w_i$ is the window coefficient multiplying the $i$th sampling in the time series and $N_{FFT}$ is the number of sampling points in fast Fourier transform~(FFT). 
$\Delta$ is the width of a frequency bin and is equal to $f_{sample}/N_{FFT}$, where $f_{sample}$ is the sampling frequency. 
The ratio $S/N$ is written in the form~\cite{Hibbs}~\cite{McNamara} 
\begin{eqnarray}
\frac{S}{N}=\frac{\alpha_1^2\eta_0^2}{8\alpha_0}\Biggl[ 1-\frac{\alpha_1^2\eta_0^2}{2\{ \alpha_0^2+(2\pi f_s)^2\} }\Biggr]^{-1}. \label{eq.SN}
\end{eqnarray}
$\alpha_0$ and $\alpha_1$ are calculated as follows, 
\begin{eqnarray}
\alpha_0&=&2W(\eta=0), \nonumber \\
\alpha_1&=&-2\frac{\partial W}{\partial \eta}(\eta=0). \label{alpha}
\end{eqnarray}
$W(t)=f(\mu+\eta)$ is the transition rate out of the stable states, where $\mu$ is a parameter related to a potential barrier and $\eta=\eta_0\cos(2\pi f_s t)$ is a stimulus modulating the potential periodically. 

\begin{figure}[tb]
\begin{center}
  \includegraphics{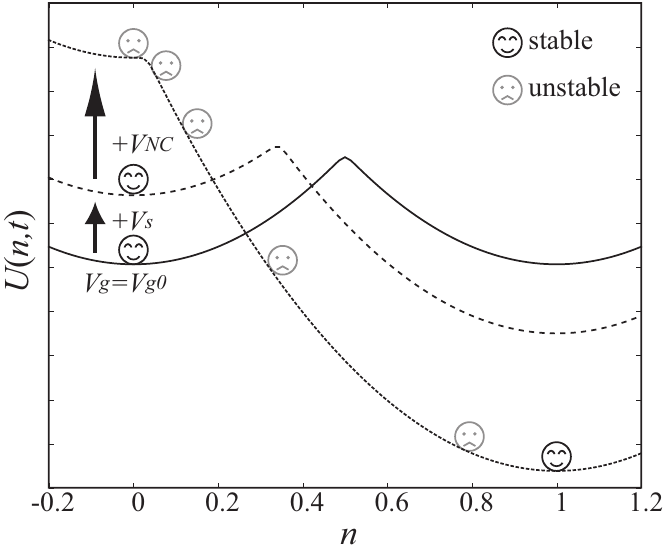}
  \caption{Bistable potential modulated by a subthreshold signal $V_s$ and critical noise $V_{NC}$. Switching of $n(t)$ can occur with the help of noise larger than critical noise. }
  \label{fig.potential}
\end{center}
\end{figure}

According to the previous results of Tsironis {\itshape et~al.\/}~\cite{Tsironis}~\cite{Rubia}, the transition rate obtained for a double-well system in the presence of colored noise is written in the form, 
\begin{eqnarray}
W(t)=\frac{V_{NC}(t)}{\tau_N(2\pi D_V)^{1/2}}\exp\Bigl( \frac{-V_{NC}^2(t)}{2D_V}\Bigr), \label{eq.W}
\end{eqnarray}
where $D_V$ is the variance of Gaussian noise $V_N(t)$, $\tau_N\equiv 1/(2f_c)$ is the correlation time of $V_N(t)$, and $\tau_N$ has to meet the ``strong color noise limit" $\tau_N\gg \tau_t$ for the validity of Eq.(\ref{eq.W}). 
$V_{NC}(t)$ is the critical noise which makes the potential barrier disappear in the presence of the signal $V_s(t)$ as shown in Fig.\ref{fig.potential}. 
$\partial U/\partial n = 0 = \partial^2 U / \partial n^2$ at $V_N(t)=V_{NC}(t)$  yields
\begin{eqnarray}
V_{NC}(t)&=&\frac{1}{2}(V_{t0}-V_{t1})\Bigl[ \{ (1+x)(1-x)\}^{1/2} - 1\Bigr] \nonumber \\
&&+ V_{t0} - V_{g0} - \frac{1}{\beta}\mathrm{sech}^{-1}x - V_s(t)\nonumber \\
&\equiv&\mu-V_s(t), \nonumber \\
x&=&\Biggl[ \frac{2}{\beta(V_{t0}-V_{t1})}\Biggr]^{1/2}. \label{eq.critical}
\end{eqnarray}
Substituting Eq.(\ref{eq.critical}) for Eq.(\ref{eq.W}) yields $W(\mu, V_{s})$ ($\eta_0$ corresponds to $V_s$'s amplitude $A_v$) and both $\alpha_0$ and $\alpha_1$ are calculated by Eq.(\ref{alpha}) and written as follows, 
\begin{eqnarray}
\alpha_0&=&\frac{2\mu}{\tau_N(2\pi D_V)^{1/2}}\exp\Bigl(\frac{-\mu^2}{2D_V}\Bigr), \nonumber \\
\alpha_1&=&\alpha_0/\mu. \label{eq.Tsironis}
\end{eqnarray}

Although we introduce $U(n, t)$ using a continuous approximation for $n$ in average meaning, the possible value of $n$ is discrete. 
Then the single electron turnstile system can not be mapped onto the double-well system exactly. 
Another different approach calculating $W$ is derived from a Fokker-Planck equation~\cite{McNamara} and $\alpha_0$ and $\alpha_1$ are calculated as follows,
\begin{eqnarray}
\alpha_0&=&\frac{1}{\tau_N(\pi)^{1/2}}\Biggl[ \int_{-\theta}^{\theta} \!\!\! e^{u^2}\phi(u)\, du\Biggr]^{-1}, \nonumber \\
\alpha_1&=&\frac{1}{\tau_N(\pi)^{1/2}}\Biggl[ \int_{-\theta}^{\theta} \!\!\! e^{u^2}\phi(u)\, du\Biggr]^{-2} \nonumber \\
&\times& e^{\theta^2}[\phi(\theta)-\phi(-\theta)], \nonumber \\
\theta&=&\gamma / (2D_V)^{1/2},  \nonumber \\
\phi(u)&=&\frac{1}{(\pi)^{1/2}}\int_{-\infty}^u \!\!\! e^{-t^2}\, dt \nonumber \\
&=&\frac{1}{2}[1+\mathrm{Erf}(u)], \label{eq.Fokker}
\end{eqnarray}
where $\mathrm{Erf}(u)$ is the error function, $\lambda$ is a parameter related to $T$ and $\gamma$ is obtained in our case as $\gamma=\lambda (V_{t0}-V_{t1})/2$. 
Note that $\eta_0$ in Eq.(\ref{eq.SN}) corresponds not to $A_v$ but to $A_v/(2D_V)^{1/2}$. 
The integral in Eq.(\ref{eq.Fokker}) can be calculated numerically. 
The strong color noise limit is also required for validity of Eq.(\ref{eq.Fokker}). 

\section{NUMERICAL SIMULATION}
We have confirmed the SR effect obeying the theory mentioned above by numerical simulation based on a Monte Carlo method~\cite{Kirihara}. 
Parameters are chosen as follows; $C=1.0$~[aF], the tunnel resistances $R_t=100$~[k$\Omega$], $C_g=0.5$~[aF], $V_{g0}=0.160218$~[V], $V_b=50$~[mV], $T=30$~[mK], $f_s=100$~[MHz], $\tau_N=125$~[ps] ($f_c=4$~[GHz]). 
In our parameters, $\tau_t$ is several tens pico seconds on average. 
The PSD related to the time series of central island voltage is computed by using a 2048 point FFT with 2~[GHz] sampling, applying a Hanning window to each segment (one segment is composed of 100-cycle waves), and averaging 100 segments. 
For $N$ in Eq.(\ref{eq.SNR}), we use an interpolated average of the PSD from neighboring 10 frequency bins. 
Averaging 100 ensembles for each noise variance $D_V$, we have obtained the SNR versus noise profile numerically. 

\begin{figure}[tb]
\begin{center}
  \includegraphics{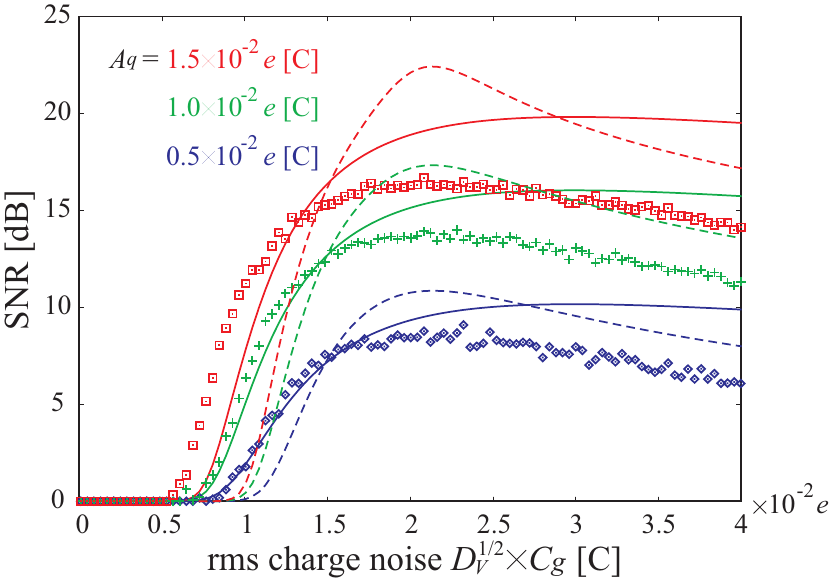}
  \caption{SNR versus noise profile. Point, solid line and dashed line plots correspond to the simulation results, the theoretical results by Eq.(\ref{eq.Tsironis}) and Eq.(\ref{eq.Fokker}), respectively. $\beta=4900$ and $\lambda=1.63$ are determined by a method of least squares.}
  \label{fig.SNR}
\end{center}
\end{figure}
Figure \ref{fig.SNR} shows three results obtained by numerical simulation (point plots), by Eq.(\ref{eq.Tsironis}) (solid line plots) and by Eq.(\ref{eq.Fokker}) (dashed line plots) for various input signal amplitude $A_q=C_gA_v$. 
$\beta=4900$ and $\lambda=1.63$ are treated as free parameters and determined by a method of least squares because it is difficult to determine them analytically. 
SR effect, wherein the SNR passes through a maximum at certain noise strength, has been confirmed from these results. 
Theoretical results obtained by Eq.(\ref{eq.Tsironis}) agree with the numerical results only in the regions wherein the SNRs increase. 
While the results obtained by Eq.(\ref{eq.Fokker}) can predict correctly the locations of the SNR maxima and the decreasing rate of the SNRs through the maxima. 
The conspicuous differences are the rise-up points and the magnitude of the SNRs especially with large $A_q$. 
This is because that the adiabatic two-state theory is valid on condition that signal amplitude is much smaller than a potential barrier. 
Another possible reason for the differences is a strong color noise limit violation because $\tau_t$ can sometimes be comparable to $\tau_N$ due to fluctuation of the tunneling time of electrons. 

\section{CONCLUSION}
We have studied stochastic resonance in a single electron turnstile both theoretically and numerically. 
These results could be applied to detection of the periodic motion of very small electric charge under noisy environment. 
One possible application is qubit detection in quantum computers ({\itshape e.g.\/}\cite{Milburn}, \cite{Tsai}) instead of using an SET (Single Electron Transistor)~\cite{Averin} or an rf-SET (radio frequency SET)~\cite{Schoelkopf}. 
Even though it detects not quantity but motion of electric charge, the superior figure, {\itshape i.e.\/} noise tolerance, is highly attractive for detecting weak signals. 

\section*{ACKNOWLEDGMENTS}
This work was supported in part by the Grant-in-Aid for Scientific Research by the Ministry of Education, Science, and Culture of Japan.

\bibliography{PLA}

\end{document}